\documentclass{aa}
\usepackage{graphicx}
\usepackage{txfonts}
\usepackage{natbib}

\renewcommand{\d}{\mathrm{d}}

\newcommand{\be}{\begin{equation}}
\newcommand{\ee}{\end{equation}}
\renewcommand{\d}{\mathrm{d}}

\begin{document} 

\title{Model-independent determination of the cosmic expansion rate. I. Application to type-Ia supernovae}

\author{Claudia Mignone\inst{1,2}\thanks{IMPRS fellow - email: \texttt{cmignone@ita.uni-heidelberg.de}}
 \and Matthias Bartelmann\inst{1}}

\institute{Zentrum f\"ur Astronomie, ITA, Universit\"at Heidelberg, Albert-\"Uberle-Str.~2, 69120 Heidelberg, Germany
  \and
  Max Planck Institut f\"ur Astronomie, K\"onigstuhl~17, 69117 Heidelberg, Germany}

\date{Received 2 November 2007 / Accepted 29 January 2008} 

\abstract 
  {}
  {In view of the substantial uncertainties regarding the possible dynamics of the dark energy, we aim at constraining the expansion rate of the universe without reference to a specific Friedmann model and its parameters.}
  {We show that cosmological observables integrating over the cosmic expansion rate can be converted into a Volterra integral equation which is known to have a unique solution in terms of a Neumann series. Expanding observables such as the luminosity distances to type-Ia supernovae into a series of orthonormal functions, the integral equation can be solved and the cosmic expansion rate recovered within the limits allowed by the accuracy of the data.}
  {We demonstrate the performance of the method applying it to synthetic data sets of increasing complexity, and to the first-year SNLS data. In particular, we show that the method is capable of reproducing a hypothetical expansion function containing a sudden transition.}
  {}


\authorrunning{C. Mignone \& M. Bartelmann}
\titlerunning{Measuring the Expansion Rate from Type-Ia Supernovae}

\maketitle

\section{Introduction}

Astronomical measurements can constrain cosmology through two functions, the cosmic expansion rate and the growth rate of cosmic structures. If space-time is on average homogeneous and isotropic and topologically simply connected, it must be described by a Robertson-Walker metric characterised by a time-dependent scale factor $a(t)$. General relativity enters when the scale factor is to be related to the energy content of the universe. However, the geometry of space-time, in particular the distance measures, are determined already once the scale factor and its first derivative are specified by the expansion function $H(a):=\dot a/a$.

The dynamics of structure growth requires more than that. Linear structure growth is commonly based on the Newtonian approximation to the continuity and Euler equations, together with Poisson's equation. They provide a valid foundation if Minkowskian space-time is a good local approximation on the scales of the structures considered. Then, the total inhomogeneous energy density drives its further evolution in time, and the relation of time to observables such as redshift again requires the expansion function $H(a)$.

We are writing this to emphasise that the expansion function is the central mathematical object underlying all cosmological constraints, augmented by the assumption of local Newtonian dynamics if structure growth in the late universe is to be included. This suggests that measurements of the expansion function itself, without any reference to Friedmann models, should be possible and much more fundamental than the common constraints of cosmological parameters entering the expansion function once the density contributions to the Friedmann models are specified.

This approach may become interesting for instance in the context of dark energy, for which our lack of understanding allows an only poorly constrained, vast range of possible models. Usually, dark-energy models are characterised by a small set of parameters and placed into the cosmic expansion rate through Friedmann's equation, substituting the conventional cosmological-constant term. While this is doubtlessly reasonable when testing specific dark-energy models, the question remains interesting what can be inferred on the cosmic expansion rate from observations without any reference to a specific model for the energy content of the universe and how it may affect cosmic dynamics.

The importance of a model-independent reconstruction of the cosmic expansion rate from luminosity distance data has been largely discussed in the literature. In \cite{STA98.1} the relations between the observational data and the expansion rate are presented, and several different techniques have been developed since then to appropriately treat the data in order to perform such a reconstruction (see,~e.g.,~\cite{HUT99.1, HUT00.1, TEG02.1, WAN05.1}). Recent reconstruction techniques with applications to data can be found in \cite{DAL03.1, DAL04.1}, where luminosity-distance data from type-Ia supernovae are combined with angular-diameter distances inferred from radio galaxies, \cite{FAY06.1}, where constraints from baryon acoustic oscillations are added to supernova data, and \cite{SHA06.1, SHA07.1}. All these methods employ a smoothing procedure in redshift bins. Also principal component analysis has been used to reconstruct the dark energy equation of state as a function of redshift \citep{HUT03.1, CRI05.1, SIM06.1}.

We are here proposing a method aiming at a direct determination of the cosmic expansion rate without reference to any specific model for the energy content of the universe. As proposed and used here, it only employs the minimal assumptions (1) that the universe is topologically simply connected, homogeneous and isotropic on average, and (2) that the expansion rate is reasonably smooth. We introduce the method in Sect.~2 and demonstrate its performance with synthetic and real type-Ia-supernova data in Sect.~3. We finish with the discussion in Sect.~4.

\section{Method}

\subsection{Preliminary remarks}

The cosmological standard model, which was defined and tightly constrained approximately during the last decade, is based on the combination of a multitude of cosmological measurements. Perhaps unduly simplifying the picture, the typical angular size of CMB temperature fluctuations constrains the overall spatial curvature and thus the sum of all energy density contributions \citep{SPE07.1}, the large-scale galaxy power spectrum \citep{TEG04.1} and the evolution of galaxy clusters \citep{ALL03.1, ALL04.1} constrain the total matter density, primordial nucleosynthesis constrains the total density of baryonic matter \citep{KNE04.1}, and difference between the total energy density and the matter density is attributed to the cosmological constant or the dark energy.

Type-Ia supernovae stand out in this context because they allow in principle to directly measure the accelerated cosmic expansion, which in the context of Friedmann models requires a dominant energy contribution with negative pressure. Measurement accuracies now seem high enough to not only favour accelerated expansion, but to exclude decelerated expansion. Besides, type-Ia supernovae constrain cosmology purely geometrically because they allow measuring how the luminosity distance depends on redshift or, equivalently, on the scale factor $a$. For recent results see \cite{AS06.1, RIE07.1, WOO07.1, DAV07.1}.

Without loss of generality, we write the expansion function $H(a)$ in the form $H(a)=H_0E(a)$, introducing an expansion function $E(a)$ arbitrarily normalised to unity at present, $a=1$. Already the Robertson-Walker metric, before its specialisation to a Friedmann metric, shows that the angular-diameter distance is given by
\begin{equation}
  D_\mathrm{A}(a)=a\,f_K[\chi(a)]\;,
\label{eq:1}
\end{equation}
with the comoving angular-diameter distance
\begin{equation}
  f_K(\chi)=\left\{\begin{array}{ll}
    \sin\chi & (K=1) \\ \chi & (K=0) \\ \sinh\chi & (K=-1)
  \end{array}\right.
\label{eq:2}
\end{equation}
and the comoving distance
\begin{equation}
  \chi(a)=\frac{c}{H_0}\int_a^1\frac{\d a'}{a'^2E(a')}\;.
\label{eq:3}
\end{equation}
Due to Etherington's relation, which holds for any space-time, the luminosity distance is
\begin{equation}
  D_\mathrm{L}(a)=\frac{f_K[\chi(a)]}{a}\;.
\label{eq:4}
\end{equation} 

In the common approach to constraining cosmological parameters and specifically the dark energy, a particular Friedmann model is adopted, whose expansion rate is written as
\begin{equation}\label{eq:erate_fried}
  E(a)=\left[
    \frac{\Omega_\mathrm{r0}}{a^4}+\frac{\Omega_\mathrm{m0}}{a^3}-\frac{\Omega_\mathrm{k0}}{a^2}+
    \Omega_\mathrm{q0}F(a)
  \right]^{1/2}
\label{eq:usual}
\end{equation} 
in terms of the present-day density parameters $\Omega_\mathrm{(r,m,k,q)0}$ for the radiation, matter, curvature and dark-energy contributions, where $\Omega_\mathrm{k0}=1-\Omega_\mathrm{r0}-\Omega_\mathrm{m0}-\Omega_\mathrm{q0}$. The function
\begin{equation}
  F(a)=\exp\left[-3\int_1^a(1+w(a'))\frac{da'}{a'}\right]
\label{eq:6}
\end{equation}
is determined by the time-dependence of the ratio $w(a)$ between the pressure and the density of the dark energy. Equations~(\ref{eq:4}) and (\ref{eq:6}) show that the observable, i.e.~the luminosity distance, is a double integral over the dark-energy equation-of-state $w(a)$, implying that its dependence on the details of $w(a)$ is quite weak. Several articles in the literature have highlighted the pitfalls which such a weak dependance produces on the possible conclusions that can be drawn on the dark-energy properties \citep{MAO01.1, MAO02.1, BAS04.1}.

\subsection{Model-independent determination of the expansion function}

Instead of specifying a particular Friedmann model and constraining the parameters contained in $E(a)$ as outlined in Eqs.~(\ref{eq:usual}) and (\ref{eq:6}), our goal is to recover the expansion rate of the universe, $E(a)$, as a function of the scale factor $a$, without assuming any specific parameterisation for it. For simplicity of notation, we put $K=0$ now, allowing us to write $f_K(w)=w$ according to Eq.~(\ref{eq:2}), which is a first-order approximation even in the case of small $K\neq0$. This simplification could be dropped if necessary without any change of principle.

Combining Eqs.~(\ref{eq:2}), (\ref{eq:3}) and (\ref{eq:4}), we write the luminosity distance as
\begin{equation}\label{eq:d_lum}
  D_\mathrm{L}(a)=\frac{c}{H_0}\frac{1}{a}\,\int_a^1\frac{\d x}{x^2E(x)}\equiv
  \frac{c}{H_0}\frac{1}{a}\,\int_a^1\frac{\d x}{x^2}e(x)\;,
\end{equation} 
where $H_0$ is the Hubble constant, and we have defined the inverse expansion rate $e(a)\equiv E^{-1}(a)$. For the sake of simplicity, we shall drop the normalising Hubble length $c/H_0$ in the following discussion, thus scaling the luminosity distance by the Hubble length.

Differentiating Eq.~(\ref{eq:d_lum}) with respect to $a$, and dropping $c/H_0$, we obtain
\begin{equation}
  D'_\mathrm{L}(a)=-\frac{1}{a^2}\int_a^1\frac{\d x}{x^2}e(x)-\frac{e(a)}{a^3}\;,
\label{eq:Dd_lum}
\end{equation} 
which can be brought into the generic form of a Volterra integral equation of the second kind for the unknown function $e(a)$,
\begin{equation}
  e(a)=-a^3D'_\mathrm{L}(a)+\lambda\int_1^a\frac{\d x}{x^2}e(x)\;,
\label{eq:volterra}
\end{equation} 
with the inhomogeneity $f(a)\equiv-a^3D'_\mathrm{L}(a)$ and the simple kernel $K(a,x)=x^{-2}$. The general parameter $\lambda$ will later be specialised to $\lambda=a$. As detailed e.g.~in \cite{ARF95.1}, Eq.~(\ref{eq:volterra}) can be solved in terms of a Neumann series,
\begin{equation}
  e(a)=\sum_{i=0}^\infty\lambda^ie_i(a)\;,
\label{eq:neu_ser}
\end{equation}
where a possible (but not mandatory) choice for the functions $e_i$ is
\begin{equation}
  e_0(a)=f(a)\;,\quad e_n(a)=\int_1^aK(a,t)e_{n-1}(t)\d t\;.
\label{eq:neu_fct}
\end{equation} 
The first guess for $e_0(a)$ is equivalent to say that the integral or the parameter $\lambda$ in Eq.~(\ref{eq:volterra}) is small. This crude approximation, which is valid in all relevant cosmological cases, is then improved iteratively until convergence is achieved.

\subsection{Application to type-Ia supernovae}

After application of the empirical relation between duration and luminosity, observations of type-Ia supernovae yield measurements of the distance moduli $\mu_{i}$ and redshifts $z_i$ for a set of $N$ objects, which can be converted into a set of luminosity distances $D_\mathrm{L}(a_i)$ dependent on the scale factors $a_i=(1+z_i)^{-1}$. For a review about type-Ia supernovae and their cosmological implications see e.g.~\cite{LEI01.1}. 

As Eq.~(\ref{eq:volterra}) shows, our analysis requires taking the derivative of the luminosity distance with respect to the scale factor. Due to measurement errors and scatter of the data about the fiducial model, it is not feasible to directly differentiate the luminosity distance data, since the result would be extremely noisy and the estimate of $D_{\mathrm{L}}'(a)$ unreliable. Thus, we need to appropriately smooth the data. We propose to do so by fitting a suitable function $D_\mathrm{L}(a)$ to the measurements $D_\mathrm{L}(a_i)$ and approximating the derivative in Eq.~(\ref{eq:volterra}) by the derivative of $D_\mathrm{L}(a)$. This choice is justified under the assumption that the derivative of the fitted data is in fact a good representation of the actual derivative of the data.

For doing this in a model-independent way, it is convenient to expand $D_\mathrm{L}(a)$ into a series of suitably chosen orthonormal functions $p_j(a)$,
\begin{equation}
  D_\mathrm{L}(a)=\sum_{j=0}^M c_jp_j(a)\;.
\label{eq:D_basis}
\end{equation}
The $M$ coefficients $c_j$ in Eq.~(\ref{eq:D_basis}) are estimated via minimisation of the $\chi^2$ function
\begin{equation}
  \chi^2=\left(\vec D_\mathrm{obs}-\bar D(\vec a)\right)^TC^{-1}
  \left(\vec D_\mathrm{obs}-\bar D(\vec a)\right)\;,
\label{eq:chi2}
\end{equation}
where $\vec D_\mathrm{obs}$ is a vector containing the $N$ measured luminosity distances, $\vec a$ is a vector of the measured scale factors, and
\begin{equation}
  \bar D(a_i)\equiv\sum_{j=0}^M c_jp_j(a_i)\equiv(P\vec c)_i
\label{eq:D_pc}
\end{equation}
is the vector of model luminosity distances to the scale factors $\vec a$.

In the final expression of Eq.~(\ref{eq:D_pc}), $P$ is an $N\times M$ matrix with elements $P_{ij}\equiv p_j(a_i)$, and $\vec c$ is the $M$-dimensional vector of expansion coefficients. Assuming that the covariance matrix $C^{-1}$ is symmetric, the set of coefficients minimising $\chi^2$ is
\begin{equation}
  \vec c=\left(P^TC^{-1}P\right)^{-1}\left(P^TC^{-1}\right)\vec D_\mathrm{obs}\;.
\label{eq:minimiz}
\end{equation}
In this representation of the data, the derivative of the luminosity-distance function is simply given by
\begin{equation}
  D_\mathrm{L}'(a)=\sum_{j=0}^M c_jp_j'(a)\;,
\label{eq:D_basis_prime}
\end{equation}
thus avoiding the noise which would be introduced by a direct differentiation of the data.

Using the linearity of the integral equation (\ref{eq:volterra}), we can now solve it for each mode $j$ of the orthonormal function set separately. Inserting the derivative of a single basis function $p'_j(a)$ in place of $D'_\mathrm{L}(a)$ in Eq.~(\ref{eq:volterra}), its contribution to the is given in terms of the Neumann series
\begin{equation}
  e^{(j)}(a)=\sum_{k=0}^\infty a^ke^{(j)}_k(a)\;,
\label{eq:neu_ser2}
\end{equation}
with
\begin{equation}
  e^{(j)}_0(a)\equiv-a^3p'_j(a)\;,\quad
  e^{(j)}_n(a)=\int_1^ae^{(j)}_{n-1}(x)x^{-2}\d x
\label{eq:neu_fct2}
\end{equation}
according to Eq.~(\ref{eq:neu_fct}). These modes of the inverse expansion function can be computed once and for all for any given orthonormal function set $\{p_j(a)\}$. Due to the linearity of the problem, the final solution is then given by
\begin{equation}
  e(a)=\sum_{j=0}^Mc_je^{(j)}(a)\;.
\label{eq:neu_sol}
\end{equation}

\subsection{Error Analysis}

We now show how the errors on the supernova distance measurements propagate into the expansion coefficients $c_j$ and eventually into the expansion rate. We consider the Fisher matrix of the $\chi^2$ function given in Eq.~(\ref{eq:chi2}),
\begin{equation}
  F_{ij}\equiv\left\langle \frac{\partial^2\chi^2}{\partial c_i\partial c_j} \right\rangle\;,
\label{eq:fisher}
\end{equation}
which is in our case given by
\begin{equation}
  F_{ij}=\sum_{k=0}^N \frac{p_i(a_k)p_j(a_k)}{\sigma^2_k}\;,
\label{eq:fisher2}
\end{equation}
where $k$ runs over all supernova measurements and the $\sigma^2_k$ are the individual errors on the luminosity distances. By the Cram\'er-Rao inequality, the errors $\Delta c_i$ satisfy
\begin{equation}
  (\Delta c_i)^2\ge(F^{-1})_{ii}\;.
\label{eq:Cramer}
\end{equation}
These errors will propagate into the estimate $e(a)$ of the (inverse) expansion function given in Eq.~(\ref{eq:neu_sol}),
\begin{equation}
  [\Delta e(a)]^2=\sum_{j=0}^M\left[\frac{\partial e(a)}{\partial c_j}\right]^2(\Delta c_j)^2=
  \sum_{j=0}^M\left[e^{(j)}(a)\right]^2(\Delta c_j)^2\;.
\label{eq:err_e}
\end{equation}
Since the expansion rate is $E(a)=1/e(a)$, its error is finally given by
\begin{equation}
  [\Delta E(a)]^2=\frac{[\Delta e(a)]^2}{e^4(a)}\;.
\label{eq:err_E}
\end{equation}

\section{Application to synthetic and real data sets}

In this section, we demonstrate using synthetic data sets how our method performs in two different model cosmologies, an Einstein-de Sitter  and a standard $\Lambda$CDM model, using simulated samples with the characteristics of both current and future surveys. A brief discussion on the convergence of the Neumann series will follow. We also present a toy model with a sudden transition in the expansion rate, and show how our method performs in this case. Application of the method to the first year SNLS data is also presented, as well as the effects of adding high redshift ($z>1$) objects.

\subsection{Illustration: Einstein-de Sitter model}\label{sect:eds}

We start with a simple and unrealistic model cosmology in order to illustrate the proposed method in detail, i.e.~an Einstein-de Sitter universe with matter-density parameter $\Omega_\mathrm{m0}=1$, vanishing cosmological constant $\Omega_\Lambda=0$ and Hubble constant $h=0.7$. The expansion function is
\begin{equation}
  E(a)=a^{-3/2}\;, \quad e(a)=a^{3/2}\;,
\label{eq:eds_E} 
\end{equation} 
and the luminosity distance is simply
\begin{equation}\label{eq:eds_D} 
D_\mathrm{L}(a)=\frac{2}{a}(1-\sqrt{a})
\end{equation}
in units of the Hubble radius $c/H_0$. A suitable choice for the orthonormal function set could start from the linearly independent set
\begin{equation}
  u_j(x)=x^{j/2-1}\;,
\label{eq:base}
\end{equation} 
which can be orthonormalised by the usual Gram-Schmidt procedure. The orthonormalisation interval should be $[a_{\mathrm{min}},1]$, where $a_{\mathrm{min}}=(1+z_\mathrm{max})^{-1}$ is the scale factor of the maximum redshift $z_\mathrm{max}$ in the supernova sample. We thus obtain a set of orthonormal functions $\{p_j(a)\}$. Projecting the distance in Eq.~(\ref{eq:eds_D}) onto the basis functions, it is straightforward to see that only the first two modes $p_0$ and $p_1$ have non-vanishing coefficients. We then use the derivatives of $p_0$ and $p_1$ to construct the corresponding Neumann series following Eq.~(\ref{eq:neu_fct2}), and from them we recover the (inverse) expansion rate. See Appendix \ref{par:app} for further detail.

\begin{figure}[ht]
\begin{center}
\includegraphics[angle=270, width=\hsize]{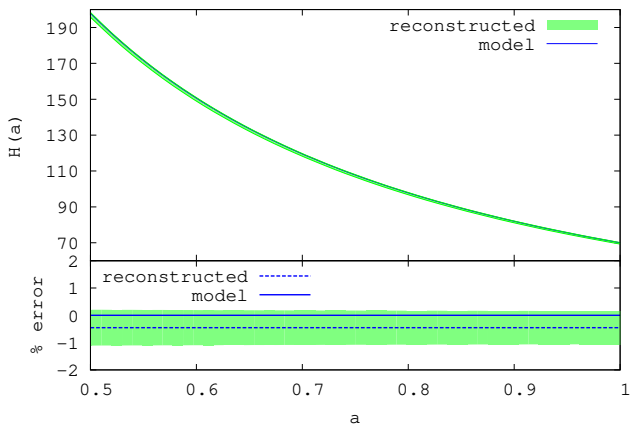}
\caption{The reconstructed expansion rate for a simulated sample of supernovae in an Einstein-de Sitter universe. The observational characteristics of the sample resemble those of the 1st year SNLS data. The green shaded area represents the reconstruction with 1-$\sigma$ errors thereof, the blue curve represents the model. The bottom plot shows the residuals between the reconstruction and the model.} \label{fig:eds}
\end{center}
\end{figure}

We applied this method to a synthetic sample of type-Ia supernovae in the Einstein-de Sitter universe. The observational characteristics of the sample, such as its size, the redshifts and the distribution of typical errors of individual measurements, are adapted to those of the first-year SNLS data \citep{AS06.1}. Thus, our sample consists of 120 supernovae up to redshift $z=1$. It enables us to determine the expansion coefficients $c_0$ and $c_1$ with relative errors of order (1-2)\%. The reconstructed expansion rate $H(a)=H_0E(a)$ is shown in Fig.\ref{fig:eds}.

The purpose of this simplified example is to show that it is possible to achieve a robust and highly accurate reconstruction of $E(a)$ when the relevant expansion coefficients can be obtained from the data with suitable significance. We now proceed to a more realistic application.

\subsection{$\Lambda$CDM model}\label{sect:la}

\begin{figure*}
\includegraphics[angle=270, width=0.33\hsize]{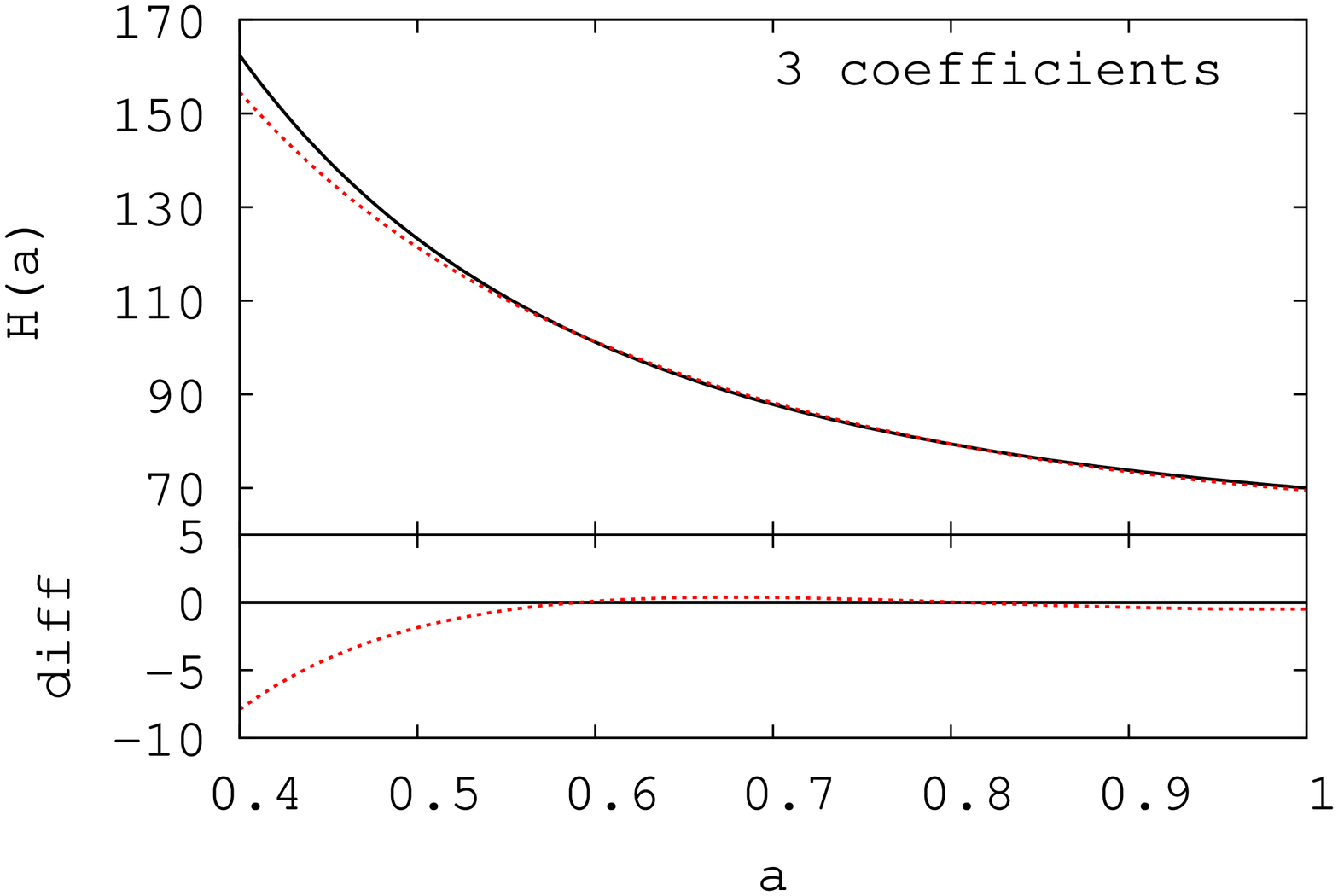}\hfill
\includegraphics[angle=270, width=0.33\hsize]{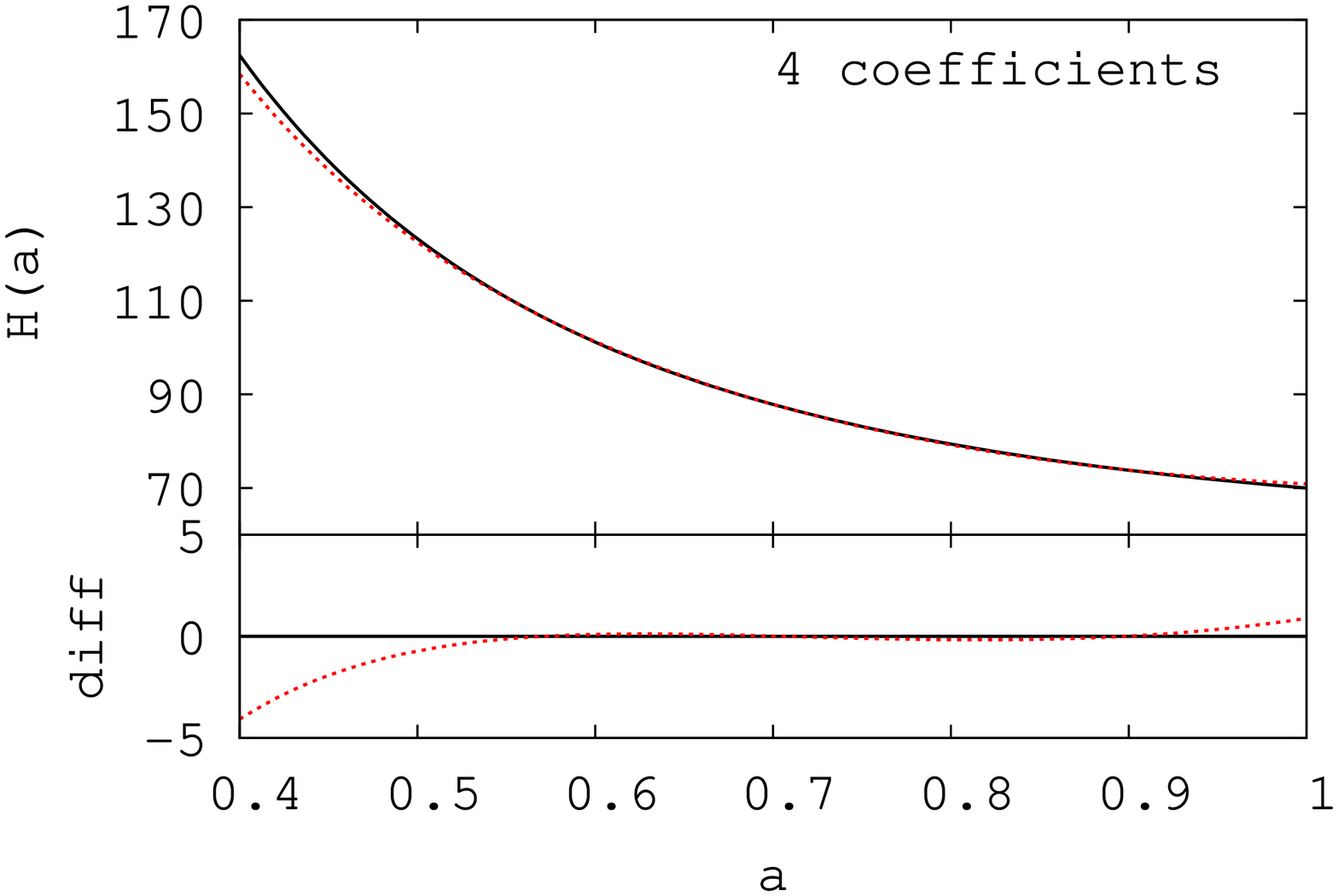}\hfill
\includegraphics[angle=270, width=0.33\hsize]{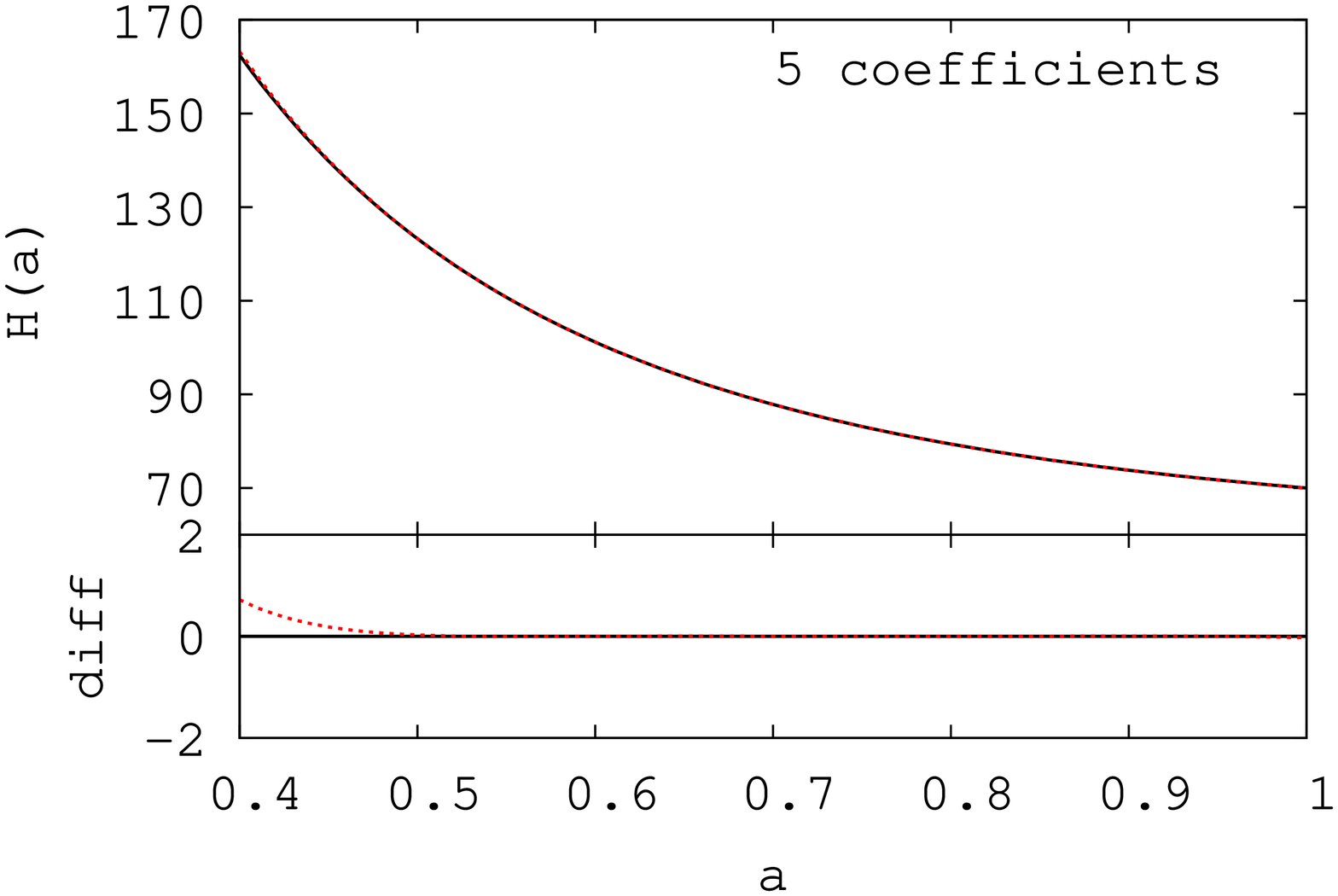}
\caption{The expansion rate in a $\Lambda$CDM model with $\Omega_\mathrm{m0}=0.3$ and $\Omega_{\Lambda}=0.7$ (solid line) and its reconstruction obtained using the \textit{true} coefficients (dashed line), truncated up to the third (\textit{left panel}), fourth (\textit{central panel}) and fifth (\textit{right panel}) coefficient, respectively. The difference between the reconstruction and the model is shown in the bottom panels. When the fifth coefficient is included, the two curves nearly coincide up to $a=0.4$.}
\label{fig:basis}
\end{figure*}

We now repeat the preceding analysis with a synthetic sample simulated in a standard $\Lambda$CDM universe with $\Omega_\mathrm{m0}=0.3$, $\Omega_{\Lambda0}=0.7$ and $h=0.7$. The expansion function is
\begin{equation}
  E(a)=\left(\Omega_\mathrm{m0} a^{-3}+\Omega_{\Lambda0} \right) ^{1/2}\;.
\label{eq:28}
\end{equation} 
In this case the first two modes of the basis $\{p_j(a)\}$ chosen before are insufficient to reproduce $D_\mathrm{L}(a)$ accurately. When we consider the \textit{true} coefficients of the expansion of $D_\mathrm{L}(a)$, which we obtain by projecting it onto the different basis functions, at least the first five coefficients are significantly different from zero. This is illustrated in Fig.~\ref{fig:basis}, where we compare the model luminosity distance to its reconstruction using the basis functions and the \textit{true} coefficient of its expansion. If only three or four coefficients are included, the reconstruction deviates significantly from the model even at low redshift.

\begin{figure}
\includegraphics[angle=270, width=\hsize]{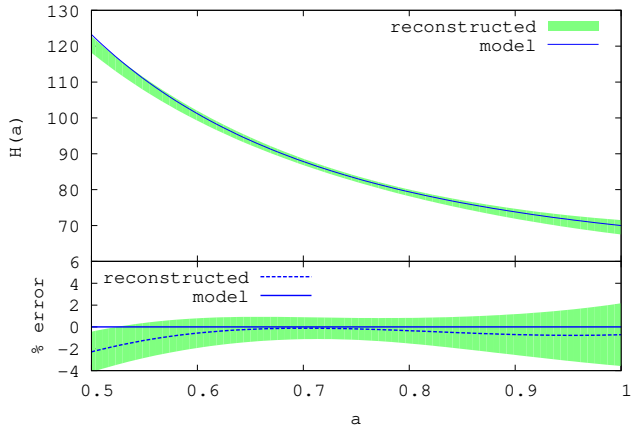}
\caption{The reconstructed expansion rate for a simulated sample of supernovae in a $\Lambda$CDM universe with observational characteristics resembling those of the 1st year SNLS data. The green shaded area represents the reconstruction with 1-$\sigma$ errors thereof, the blue curve represents the model. We made use of three coefficients. The bottom plots show the residuals between the reconstruction and the model.}
\label{fig:la}
\end{figure}

However, the measurement errors on the data play a crucial role in this analysis. With current standard data sets, only the first three coefficients can be determined significantly, while more than just three are needed to achieve an accurate reconstruction with the proposed basis functions. We show the reconstructed expansion function, obtained including three coefficients, in Fig.~\ref{fig:la}, where it is compared to the expansion rate of the underlying cosmological model. If we consider only the first three modes, \textit{all} the coefficients are statistically significant, although we know from the theoretical model that the reconstruction is incomplete. The errors on the first two coefficients $c_0$ and $c_1$ are of order (1-2)\%, increasing to 8\% on $c_2$. The errors on higher-order coefficients are larger than the coefficients themselves, indicating that they become compatible with zero and should therefore be excluded from the reconstruction.

The precision with which coefficients can be determined from the data is likely to improve dramatically with future generation, space-based supernova surveys such as the Supernova/Acceleration Probe (SNAP, \cite{ALD04.1}). SNAP is expected to measure high-quality light curves and spectra for $\approx2000$ type-Ia supernovae in the redshift range $0.1<z<1.7$. With data of such high quality it will become possible to achieve an extremely accurate reconstruction of the expansion rate with our method. As discussed above, we need at least five coefficients in order to reconstruct the expansion rate of an underlying $\Lambda$CDM model with the set of functions described above.

\begin{figure}
\includegraphics[angle=270, width=\hsize]{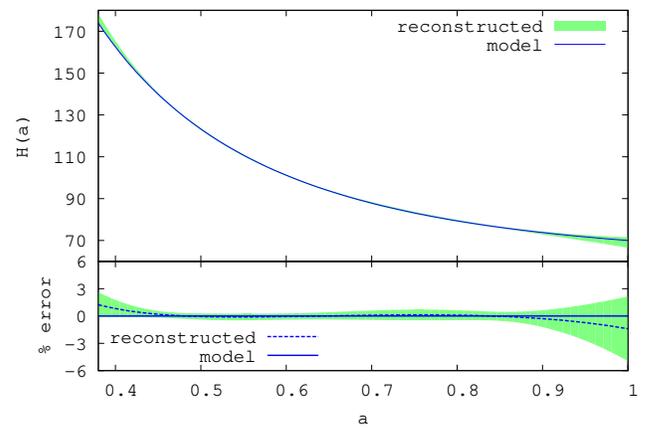}
\caption{The reconstructed expansion rate for a simulated sample of supernovae in a $\Lambda$CDM universe with the forecast observational characteristics of the SNAP experiment. The green shaded area represents the reconstruction with 1-$\sigma$ errors thereof, the blue curve represents the model. We made use of five coefficients. The bottom plot shows the residuals between the reconstruction and the model.}
\label{fig:snap}
\end{figure}

We produced a synthetic data set with SNAP characteristics, following the expected SNAP redshift distribution reported in \cite{SHA06.1}. We also added 25 more supernovae with $z<0.1$, as done in \cite{SHA06.1}, which are supposed to be observed by future low-redshift supernova experiments. Applying our reconstruction technique, we significantly constrain the first five coefficients, with errors on the first two coefficients being of order 0.1\%. The result, obtained using five coefficients, is shown in Fig.~\ref{fig:snap} together with 1-$\sigma$ errors.

The choice of the orthonormal function set is in general arbitrary. Obviously, for each underlying model there will be a preferred function set, in the sense that the number of coefficients required to reproduce the expansion rate is minimal when using such a set. It is certainly possible to find a more suitable function set for the $\Lambda$CDM model, but since our ultimate goal is to reconstruct the expansion rate from the observed data introducing as little theoretical prejudice as possible, we are not primarily interested in finding the most suitable function set to reproduce the $\Lambda$CDM  expansion rate. Thus, we only made use of the basis described in Sect.~\ref{sect:eds}.

\subsection{Convergence of the Neumann series}

A separate, but related issue is to what power of the parameter $\lambda$ we need to follow the Neumann series, or, equivalently, to what power $k$ of the scale factor $a$ we need to expand in Eq.~(\ref{eq:neu_ser2}). The truncation criterion must again be based on the quality of the data. Convergence of the series is achieved at different powers $k$ for different redshift intervals. In order to achieve convergence on the interval $0.5\le a\le1$, the series can be truncated after $k=4$. However, the inclusion of a fourth-order term produces a difference to the preceding three orders which is already within the error bars, and can therefore be neglected. This trend is clearly enhanced when more coefficients are included in the reconstruction, since in this case the errors are larger. 

\subsection{Recovery of sudden transitions in the expansion rate}

As we have emphasised above, our method can obtain the expansion function $E(a)$, or rather its reciprocal $e(a)$, based on a representation of the derivative of the measured data. We argue that dealing with the derivative of luminosity distance data is not expected to cause a major problem, based on the reasonable assumption that the luminosity distance is a very smooth function. As it is evident from Eq.~(\ref{eq:d_lum}), $D_\mathrm{L}(a)$ is related to the expansion function via an integral. Hence, even if $E(a)$ had a peculiar feature at some intermediate redshift, this would be smoothed out by the integration.

We address the issue by means of a toy model where the expansion function has indeed a sudden transition. We construct the toy model starting from the expansion rate of the Einstein-de Sitter model and deforming it by a gentle jump at some intermediate value $a_*$ of the scale factor,
\begin{equation}
  E(a)=\left\{\begin{array}{ll}
  -\arctan\left[ \gamma\left( a-a_*\right)\right] +\delta & \quad(a>\tilde{a}) \\
  a^{-3/2}+1 & \quad(a\leq\tilde{a}) \;.
  \end{array}\right.
\label{eq:irreg_E}
\end{equation} 

\begin{figure}
\includegraphics[angle=270, width=\hsize]{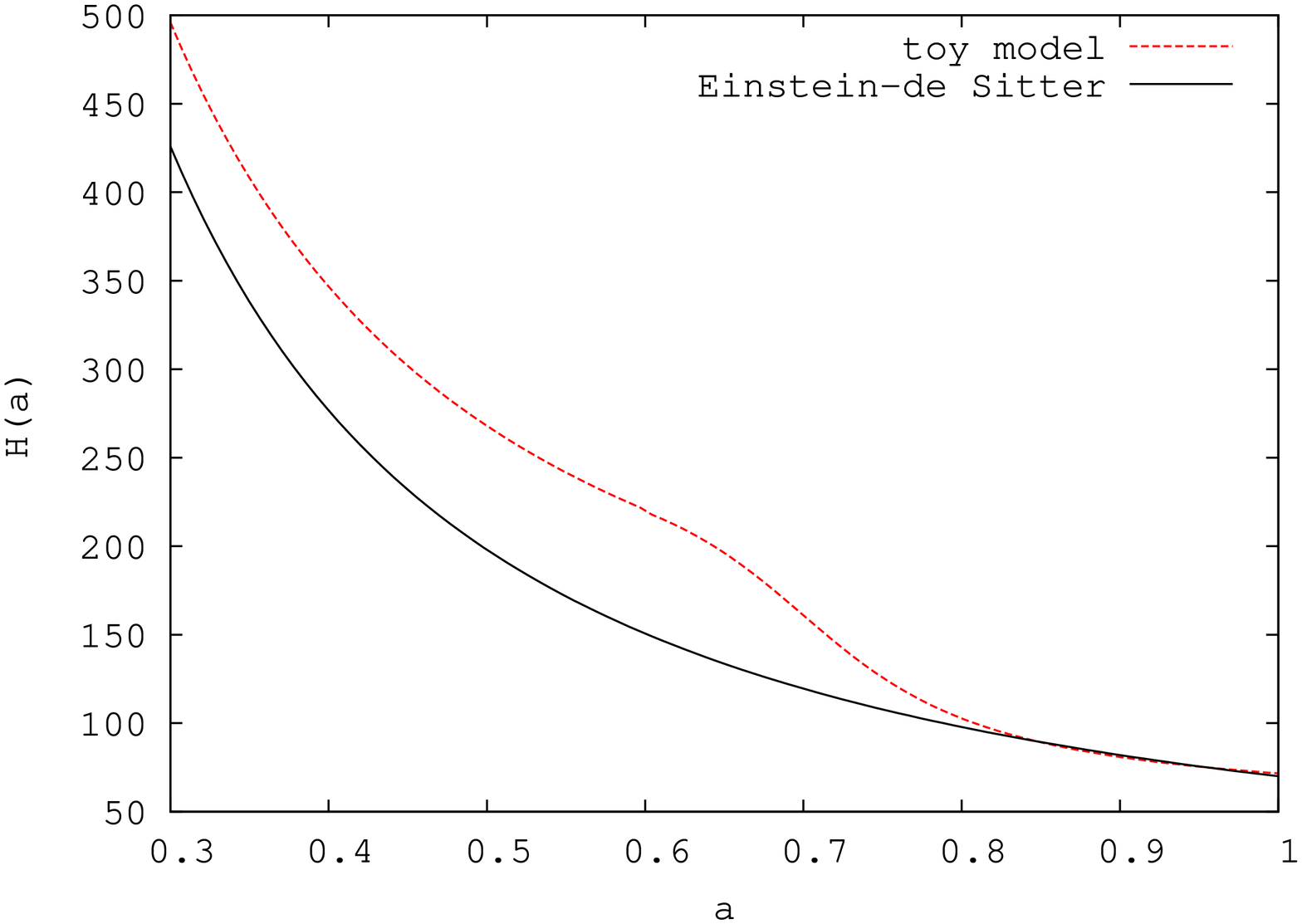}
\includegraphics[angle=270, width=\hsize]{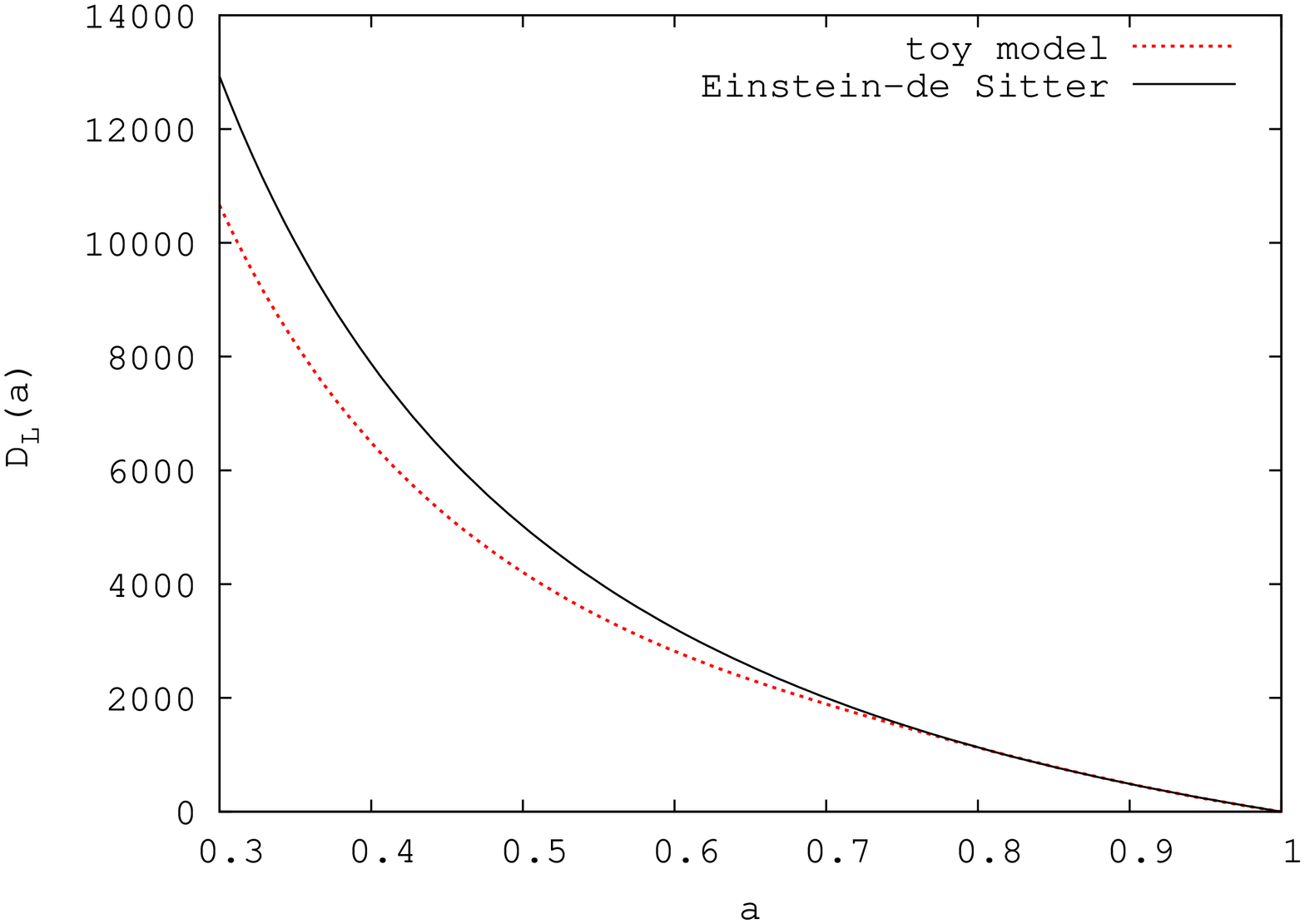}
\caption{The expansion rate of our toy model compared to the Einstein-de Sitter one (\textit{upper panel}), and the corresponding luminosity distance (\textit{lower panel}). The parameters for the toy model are: $a_*=0.7$, $\tilde{a}=0.6$, $\gamma=11$, $\delta=2.3$.} 
\label{fig:toy}
\end{figure}

Using Eq.~(\ref{eq:d_lum}), we can obtain the corresponding luminosity distance, which is quite smooth and deviates from its Einstein-de Sitter counterpart in a way depending on $a_*$. The expansion rate and the luminosity distance of this toy model are plotted in Fig.~\ref{fig:toy}, compared to those of the Einstein-de Sitter model.

\begin{figure}
\includegraphics[angle=270, width=\hsize]{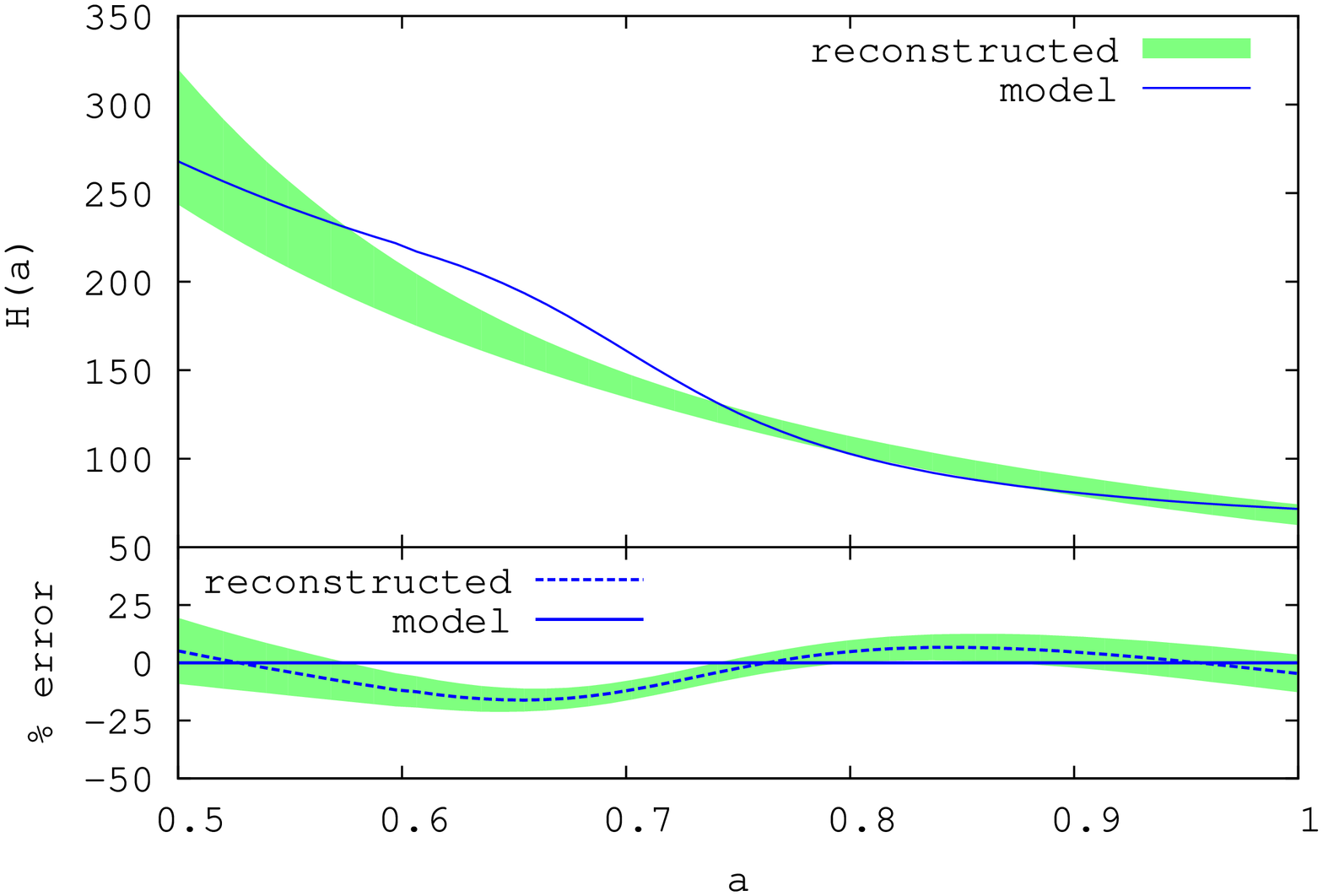}
\includegraphics[angle=270, width=\hsize]{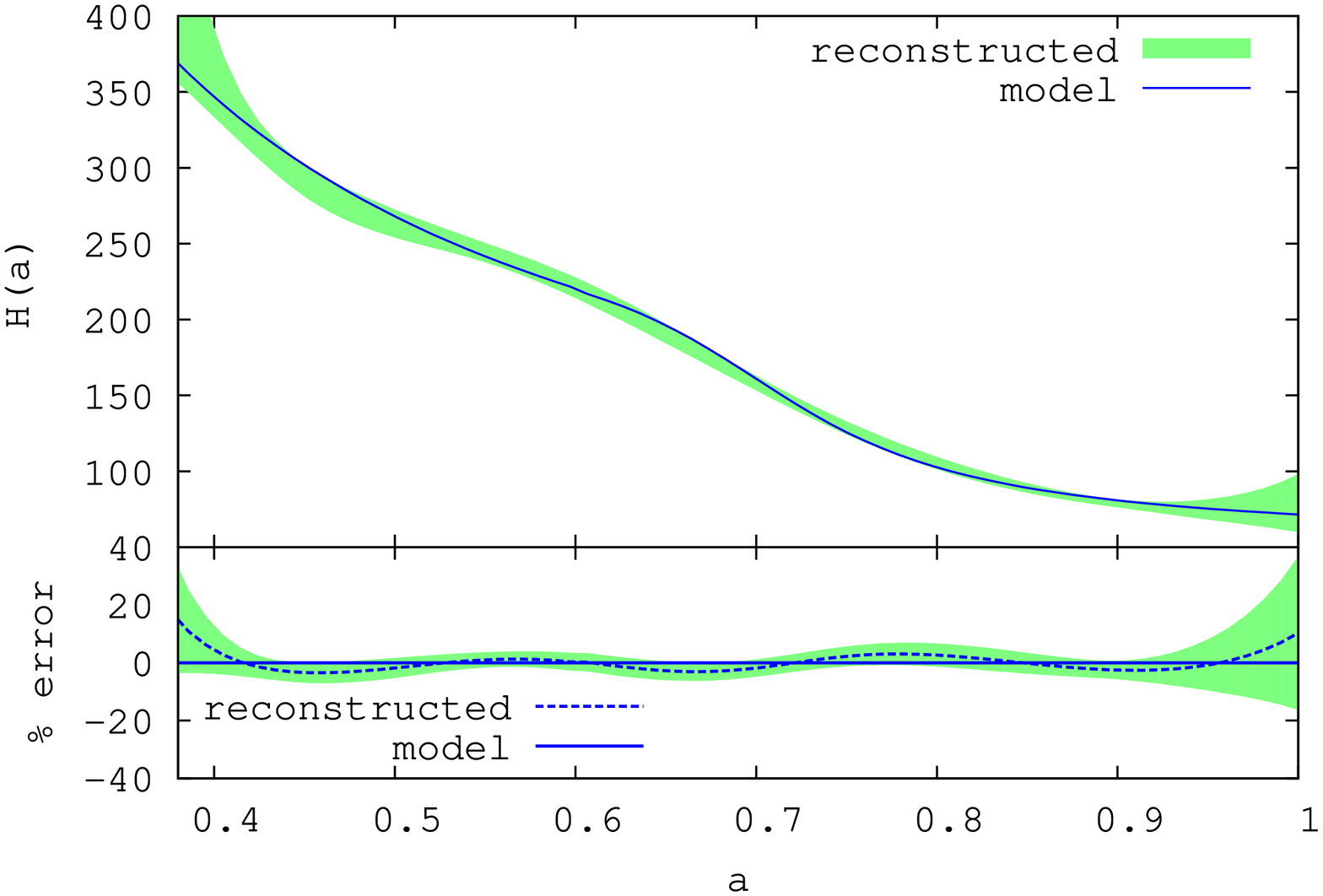}
\caption{The expansion rate of our toy model (blue curve) and its reconstruction, with 3-$\sigma$ errors thereof (green shaded area), obtained from a SNLS-like data set with three coefficients (\textit{upper panel}), and from a SNAP-like data set with six coefficients (\textit{lower panel}). The bottom plots show the residuals between the reconstruction and the model.} 
\label{fig:smooth_rec}
\end{figure}

Again, we create a synthetic sample of type-Ia supernovae within this model, with the same observational characteristics of either SNLS or SNAP, and we apply our reconstruction procedure. In order to reproduce the transition feature, we need more than three coefficients. This is not feasible with SNLS-like data because coefficients beyond the third lose significance. With a  SNAP-like sample instead, the expansion rate can be recovered. The results obtained with both synthetic samples, with three coefficients for the SNLS and six for the SNAP case, are shown in Fig.~\ref{fig:smooth_rec} together with their 3-$\sigma$ errors. Figure~\ref{fig:smooth_rec} shows that our method can also recover expansion histories with unexpected transitions, even though the reconstruction is less accurate than that of a perfectly smooth expansion rate.

\begin{figure}
\includegraphics[angle=270, width=\hsize]{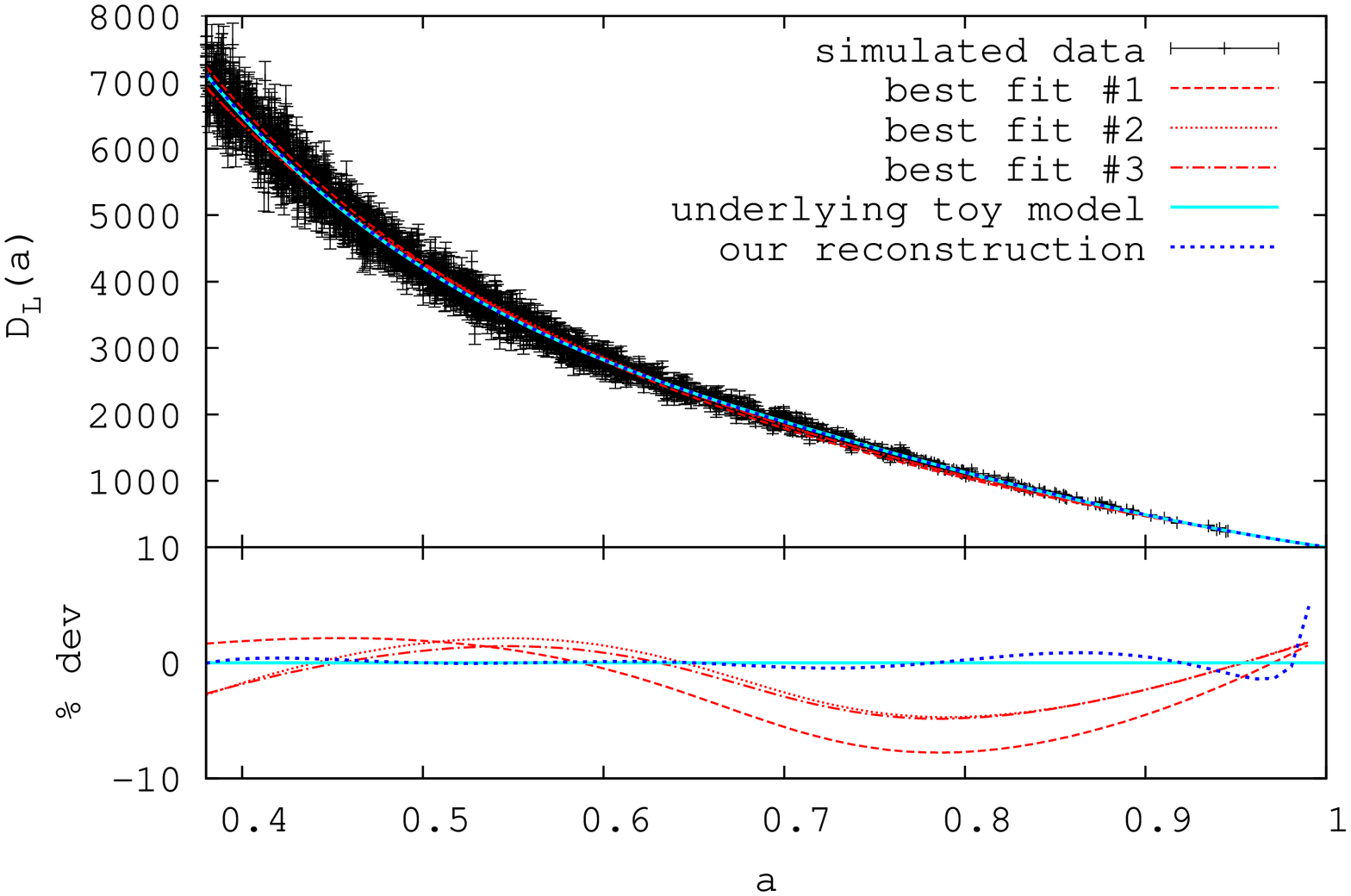}
\includegraphics[angle=270, width=\hsize]{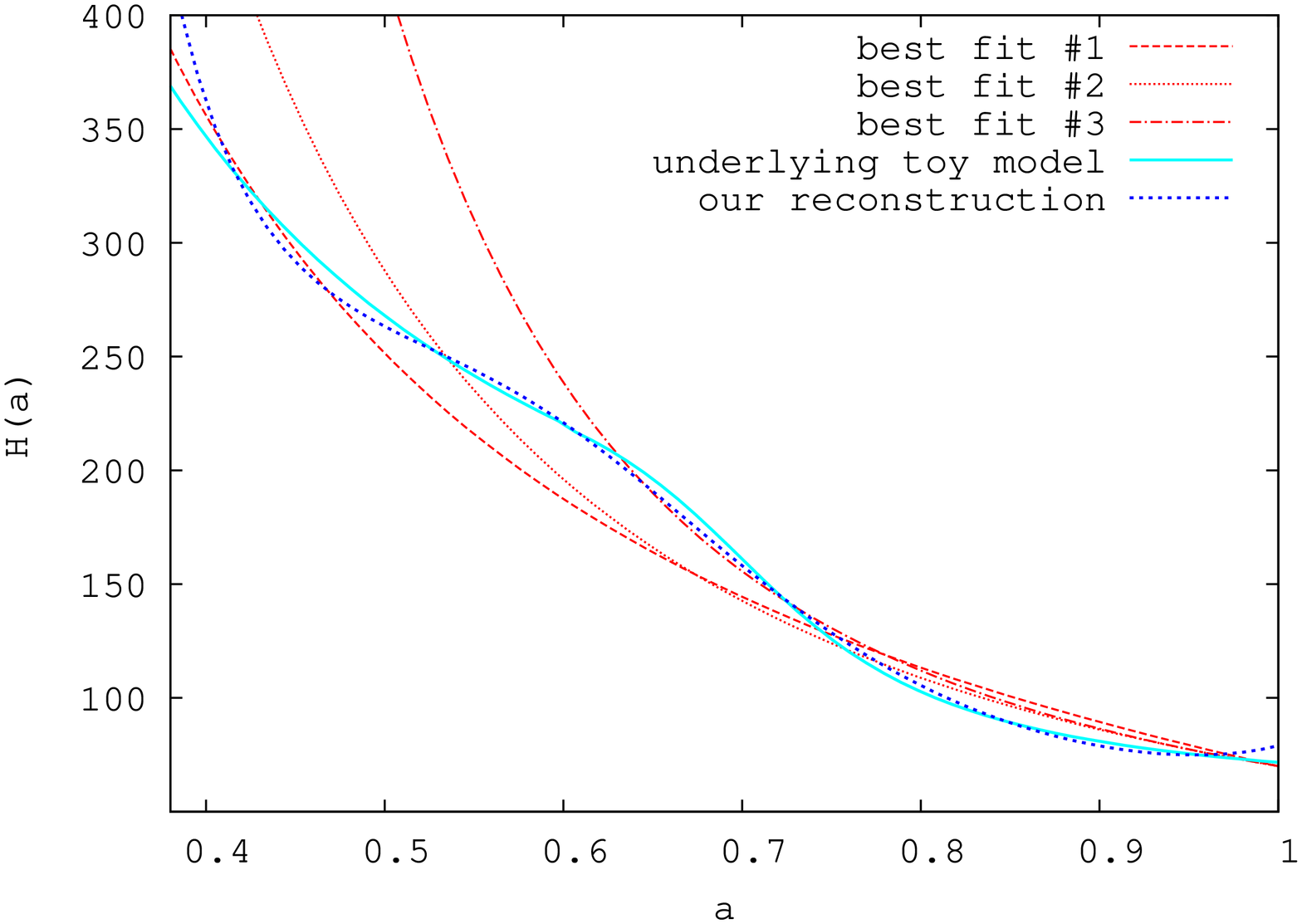}
\caption{\textit{Upper panel:} The luminosity distance of the toy model (cyan curve) together with the SNAP-like simulated sample (black points), compared to our fit (blue dashed curve) and three other cosmological fits (red curves). The bottom plot shows the residuals between the different fits and the model. \textit{Lower panel:} The expansion rate of the toy model, of our reconstruction and of the other models. The different red curves correspond to three different fits to a flat $\Lambda$CDM: in case \#1 (red dashed curve) we impose $w=-1$ and let only $\Omega_\mathrm{m0}$ vary, in case \#2 (red dotted curve) we let both $\Omega_\mathrm{m0}$ and $w$ (constant in redshift) vary, and in model \#3 we let $\Omega_\mathrm{m0}$, $w_0$ and $w_a$ vary, according to Eq.~(\ref{eq:linder}). } 
\label{fig:cf_fits}
\end{figure}

We also try to fit this sample to a flat $\Lambda$CDM model and explore the parameter space spanned by $\Omega_\mathrm{m0}$ and $w$. The dark-energy equation-of-state parameter $w$ is allowed to differ from $-1$, first assuming that it is constant in redshift and then parameterising its time evolution according to
\begin{equation}\label{eq:linder}
  w(a)=w_0+w_a(1-a)\;,
\end{equation}
as proposed by \cite{CHE01.1} and \cite{LIN03.1}. We find that all the models we considered are capable of producing good fits to the luminosity distance data, but they all fail to reproduce the underlying expansion rate when the best fit parameters are inserted back into Eq.~(\ref{eq:erate_fried}). In most cases, the likelihood has more than one maximum, since different combinations of the considered parameters constrain the two different branches of the expansion rate. Unless the time evolution of the dark-energy equation-of-state is modelled \textit{ad hoc}, it is very unlikely to reproduce the sudden feature of the toy model in this way. However, our method achieves this because the parameters involved in our fit trace the relation the between luminosity distance and the expansion rate. The different results obtained with the usual approach and our method for the fit to the luminosity distance and for the expansion rate are displayed in Fig.~(\ref{fig:cf_fits}).

\subsection{Application to the first-year SNLS data}

We finally apply our method to the first-year SNLS data \citep{AS06.1}. The sample consists of 71 new supernovae observed from the ground with the Canada-France-Hawaii Telescope, the farthest being at redshift $z=1.01$, plus 44 nearby supernovae taken from the literature. Thus, the total sample contains 115 supernovae in the redshift range $0.015<z<1.01$.

Assuming a flat, $\Lambda$CDM universe with constant $w=-1$, \cite{AS06.1} obtained a best fit of $\Omega_\mathrm{m0}=0.263\pm0.037$. Releasing the flatness assumption and adding constraints from the baryon acoustic oscillations (BAO) measured in the SDSS \citep{EIS05.1}, they obtained $\Omega_\mathrm{m0}=0.271\pm0.020$ and $\Omega_{\Lambda0}=0.751\pm0.082$. Furthermore, they investigated models with constant equation of state $w\neq-1$: assuming flatness and the BAO constraints, their best-fit parameters are $\Omega_\mathrm{m0}~=~0.271\pm~0.021$ and $w=-1.023\pm0.087$.

\begin{figure}
\includegraphics[angle=270, width =\hsize]{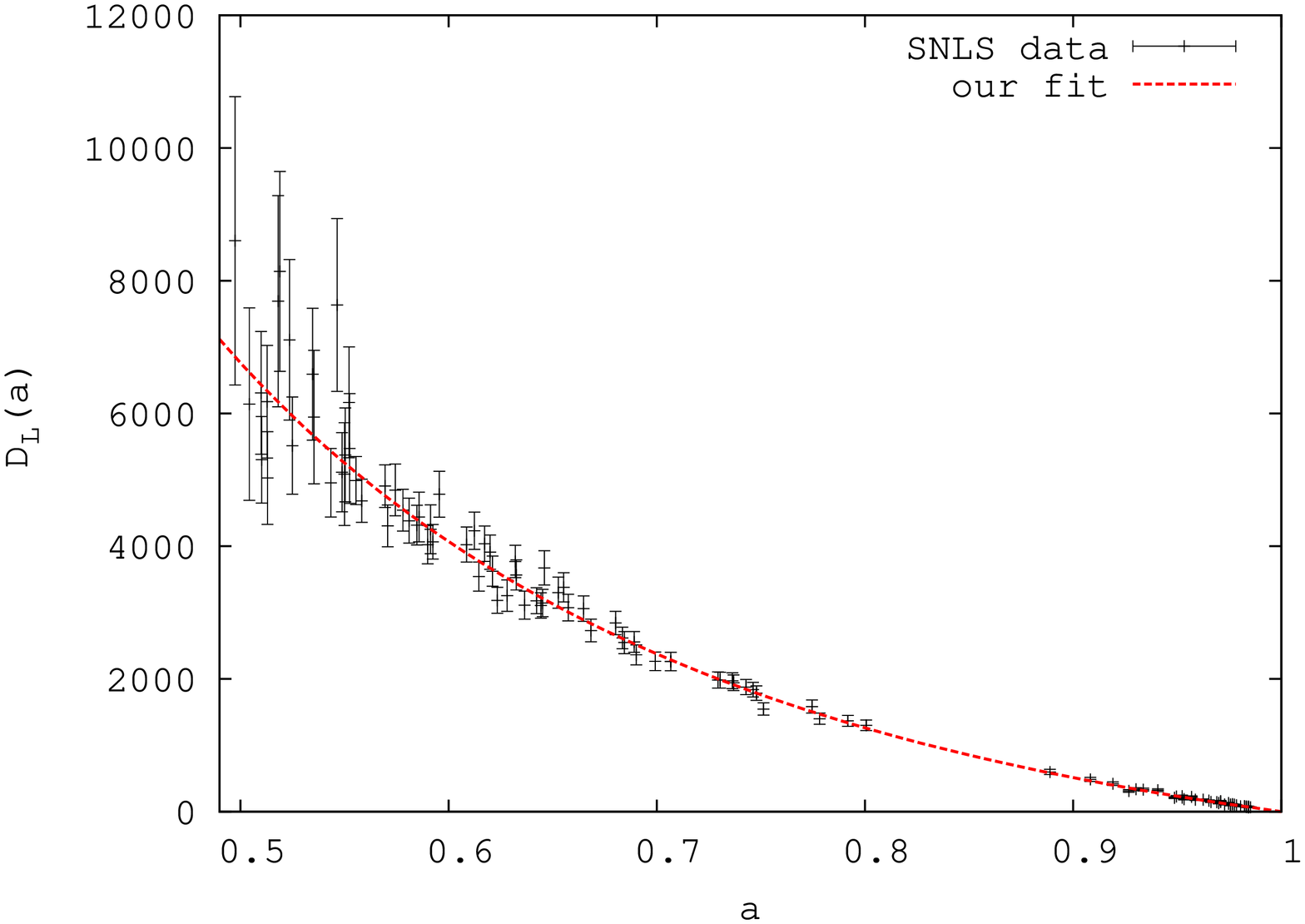}\\
\includegraphics[angle=270, width =\hsize]{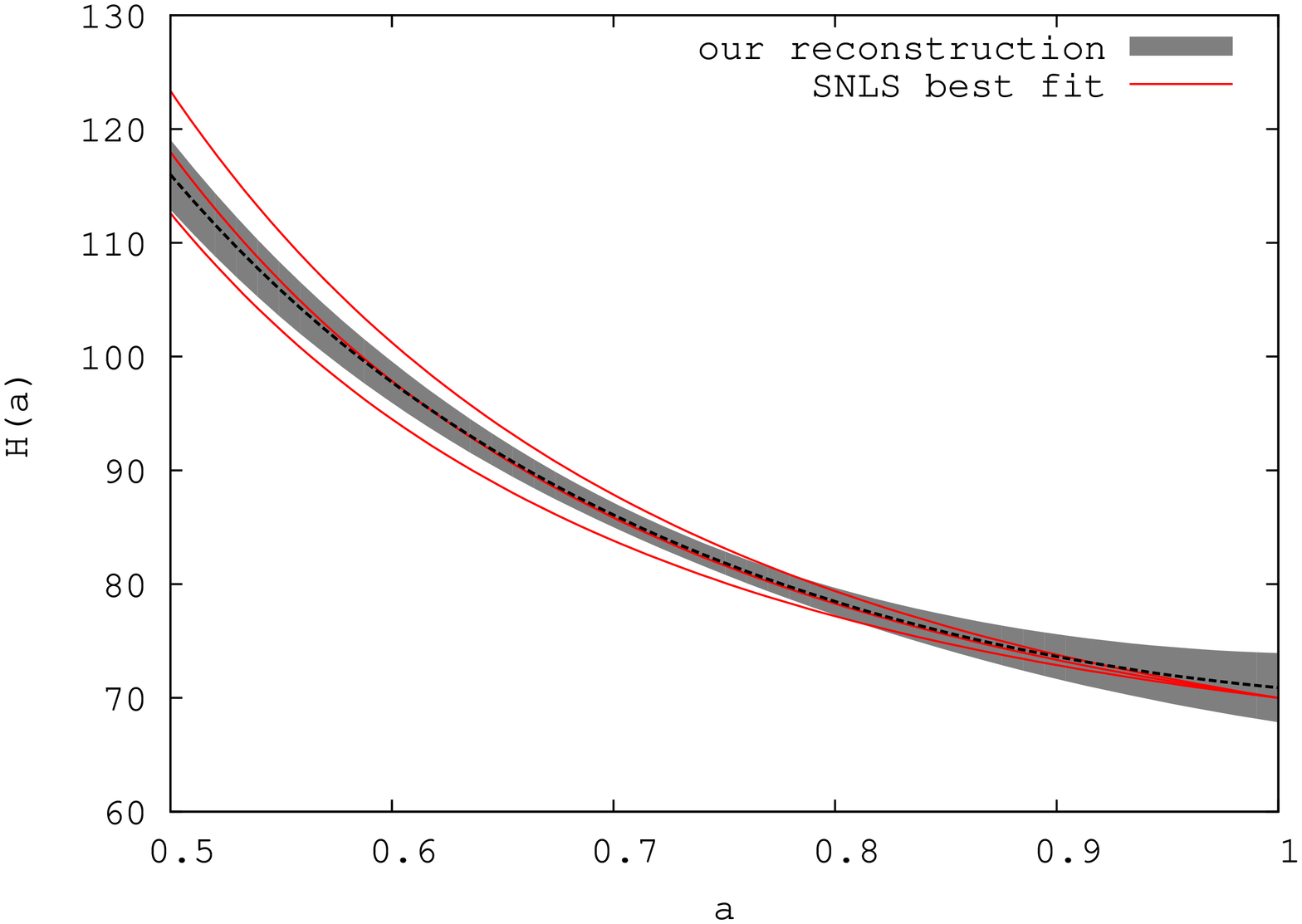}
\caption{\textit{Upper panel:} The 1st year SNLS sample and our fit for the luminosity distance. \textit{Lower panel:} Our reconstruction of the expansion rate and 1-$\sigma$ errors thereof (gray shaded area), compared to the expansion rate of a $\Lambda$CDM model with the best fit parameter from \cite{AS06.1}, $\Omega_\mathrm{m0}=0.263\pm0.037$, and 1-$\sigma$ errors thereof (red curves).}
\label{fig:snls}
\end{figure}

The fit to the luminosity-distance data obtained applying our method to this sample, with the orthonormal function set described in Sect.~\ref{sect:eds}, is shown in the upper panel of Fig.~\ref{fig:snls}. It yields three significant expansion coefficients because the data quality especially at high redshift does not allow constraints of higher-order modes, as discussed in Sect.~\ref{sect:la}.

The expansion rate reconstructed with our method is compared in the lower panel of Fig.~\ref{fig:snls} to that of the best-fit model of the SNLS analysis, i.e. a flat $\Lambda$CDM with $\Omega_{\mathrm{m0}}=0.263$.

\subsection{Extending the sample beyond z=1}

Another interesting problem concerns what could improve the performance of the method. Clearly, both a larger sample of supernovae and a better accuracy in the individual measurements would help reducing the errors on the coefficients, which would eventually enable a significant estimate of more coefficients, and thus a more precise reconstruction of the expansion rate. From a mathematical point of view, adding more objects and reducing the uncertainties are equivalent: a sample four times larger than another yields the same results obtained with the smaller sample, if its error bars were to be reduced by one half. However, since the measurement accuracy cannot be indefinitely shrunk below a given limit, because of systematic uncertainties, a long run strategy to make best use of the method would be to increase the size of the sample.

Extending the sample to higher redshift can also help reducing the errors on the estimated coefficients. We investigated the issue by means of an extremely simplified example: we added to our previously described simulated $\Lambda$CDM SNLS-like sample of supernovae up to $z=1$ an additional set of 20 objects, uniformly distributed between $z=1$ and $z=1.7$. The excess in the Fisher matrix due to the inclusion of the higher-z sample influences the errors on the coefficients, reducing the first by $\sim10\%$ and the following ones by $\sim20\%$, whereas if we only added 20 more objects with $z\leq1$ the gain would be smaller. However, this still does not allow more than three coefficients to be significantly pinned down, with the orthonormal function set described in Sect.~\ref{sect:eds}.

We also apply our method to a supernova sample which extends beyond $z=1$, namely the one compiled by \cite{DAV07.1}, including the combined ESSENCE/SNLS/nearby dataset from \cite{WOO07.1} and the HST data from \cite{RIE07.1}. It contains 192 supernovae, of which 15 with $1<z<1.75$. Although this sample contains more objects than the SNLS and extends to higher redshifts, the quality of the reconstruction achieved is not better than the one obtained using the SNLS data set. In fact, the errors on the coefficients are slightly larger, because the individual uncertainties on the distance moduli in the extended ESSENCE sample are significantly higher than the SNLS ones (at least for $z<0.8$), due to the different way luminosity distances are estimated from the photometric data by the two groups (a review on the different supernova light-curve fitters is given in \cite{CON07.1}).

\section{Discussion}

We are proposing a method to constrain the expansion function of the universe without assuming any specific model for Friedmann-type expansion. If the universe is isotropic, homogeneous and simply connected, it is described by a Robertson-Walker metric. Cosmological measurements can generally only constrain one of two functions of time, the expansion rate and the growth of structures. We have shown here how the expansion function can be observationally constrained without reference to specific assumptions on the time evolution of the terms in Friedmann's equation and their parameterisation in terms of density parameters. The issue may become important in search of constraints for a dynamical dark energy component, whose behaviour is so far only very poorly known. Since it is unclear how its energy density contribution to the cosmic fluid may change in time, any guessed parameterisation may be erroneous, hence a parameter-free recovery of the cosmic expansion rate may turn out advantageous.

We demonstrate our method here using the luminosity-distance measurements obtained from type-Ia supernovae as a model. Since the luminosity distance is a cosmological observable depending on space-time geometry only, the dynamics of structure growth does not enter yet. The method proceeds in two essential steps. First, the integral relation between the expansion function and the luminosity distance is transformed into a Volterra integral equation of the second kind. Under the relevant conditions, its solutions are known to exist and to be uniquely described in terms of a convergent Neumann series. In other words, the method is guaranteed to return \textit{the} unique expansion rate of the universe within the accuracy limits allowed by the data.

The drawback of the transformation to a Volterra integral equation is that the derivative of the luminosity distance with respect to the scale factor is needed to start the Neumann series. Derivatives of data are notoriously noisy and should be avoided. We propose to expand the luminosity distance into an initially arbitrary orthonormal function set, fit its expansion coefficients to the data and then use the derivative of the series expansion instead of the derivative of the data. Suitable orthornormal function sets can be constructed by Gram-Schmidt orthonormalisation from any linearly independent function set. The only condition so far is that the number of coefficients required to fit the data should be minimal.

Once the orthonormal function set is specified, the Neumann series can be constructed beforehand for all its members. The measured coefficients of the series expansion directly translate to the solution for the expansion function. The convergence criterion for the Neumann series is determined by the data quality, as is the number of orthonormal modes in the series expansion of the data.

Applications to synthetic data samples of increasing complexity are very promising. In particular, we showed that an expansion function containing a sudden transition can be faithfully recovered by our method provided sufficient quality of the input data. We also apply our method to the first-year SNLS data and show the expansion function recovered, and notice how the inclusion of higher-redshift ($z>1$) objects could improve the performance of the method.

In future studies, we shall investigate how our method can be used to determine the single expansion function underlying all cosmological measurements used.

\acknowledgements{It is a pleasure to thank M.~Maturi for carefully reading the manuscript and providing helpful suggestions. We also wish to thank L.~Moscardini, M.~Meneghetti, B.~M\'enard, H.~W.~Rix, C.~Baccigalupi and P.~Astier  for useful comments and suggestions. We also thank D.~Huterer, Y.~Wang and D.~Polarski for pointing out some useful remarks, and an anonymous referee for comments which helped us improving the paper. CM acknowledges E.~Ziegler and P.~Melchior for computational advice, and C.~Fedeli for helpful discussions. CM is supported by the International Max Planck Research School (IMPRS) for Astronomy and Cosmic Physics at the University of Heidelberg.}

\appendix

\section{Exact Solution for the Einstein-de Sitter case}\label{par:app}

Here we show how to construct the (inverse) expansion rate of the Einstein-de Sitter model from the first two modes of the function set obtained applying Gram-Schmidt orthonormalisation to the set of linearly independent functions specified by Eq.~(\ref{eq:base}). The first two modes are
\begin{equation}\label{eq:p0p1} 
p_0(x)=\frac{1}{\sqrt{\alpha}}\frac{1}{x}\;, \quad 
p_1(x)=\frac{1}{\sqrt{C}}\left( \frac{1}{\sqrt{x}}+\frac{2\beta}{x}\right) \;,
\end{equation}
with
\begin{eqnarray}\label{eq:const} 
\alpha&=&\frac{1-a_\mathrm{min}}{a_\mathrm{min}}\;,\nonumber\\  C&=&4-\frac{8}{1-\sqrt{a_\mathrm{min}}}-\ln a_\mathrm{min}\;, \nonumber\\ 
\beta&=&\frac{1}{\sqrt{\alpha}}\frac{-1+\sqrt{a_\mathrm{min}}}{\sqrt{1-a_\mathrm{min}}}\;.
\end{eqnarray} 
It is straightforward to see, by projecting the distance in Eq.~(\ref{eq:eds_D}) onto the basis functions, that only the first two modes are needed, i.e.
\begin{equation}
  D_\mathrm{L}(a)=\sum_{j=0}^1 \tilde{c}_jp_j(a)\;,
\label{eq:D_basis_eds}
\end{equation} 
where $\tilde{c}_j=\int_{a_\mathrm{min}}^1D_\mathrm{L}(a)p_j(a)da$ stands for the $j$-th \textit{true} coefficient of the expansion. In this case $\tilde{c}_0=2(1+2\beta)\sqrt{\alpha}$ and $\tilde{c}_1=-2\sqrt{C}$.

From the derivative of $p_0(a)$,
\begin{equation}
  p'_0(x)=-\frac{1}{\sqrt{\alpha}}\frac{1}{x^2}\;,
\end{equation} 
we can construct the zero-th order Neumann series following Eq.~(\ref{eq:neu_fct}):
\begin{eqnarray}\label{eq:0fct}
e^{(0)}_0(a)&=&-a^3p'_0(a)=\frac{1}{\sqrt{\alpha}}a\;,\nonumber\\
e^{(0)}_1(a)&=&\int_1^a\frac{\d x}{x^2}e^{(0)}_0(x)=\frac{1}{\sqrt{\alpha}}\ln a\;,\nonumber\\
e^{(0)}_2(a)&=&\int_1^a\frac{\d x}{x^2}e^{(0)}_1(x)=\frac{1}{\sqrt{\alpha}}\frac{a-1-\ln a}{a}\;.
\end{eqnarray}
Up to second order, the zero-th order Neumann series for the (inverse) expansion rate is
\begin{eqnarray}\label{eq:0ser}
e^{(0)}(a)&=&\sum_{k=0}^2 a^k e_k^{(0)}(a)\\&=&
\frac{1}{\sqrt{\alpha}}\left(a+a\ln a+a(a-1)-a\ln a\right)=\frac{1}{\sqrt{\alpha}}a^2\;.
\end{eqnarray} 
Again, from the derivative of $p_1(a)$
\begin{equation}
p'_1(x)=-\frac{1}{\sqrt{C}}\left( \frac{1}{2x^{3/2}}+\frac{2\beta}{x^2}\right)\;,
\end{equation}
we can construct the first-order Neumann series:
\begin{eqnarray}\label{eq:1fct}
e^{(1)}_0(a)&=&-a^3p'_1(a)=\frac{1}{\sqrt{C}}\left( \frac{a^{3/2}}{2}+2\beta a \right) \;,\nonumber\\
e^{(1)}_1(a)&=&\int_1^a\frac{\d x}{x^2}e^{(1)}_0(x)=\frac{1}{\sqrt{C}}
    \left( \sqrt a+2\beta\ln a-1\right) \;,\nonumber\\
e^{(1)}_2(a)&=&\int_1^a\frac{\d x}{x^2}e^{(1)}_1(x)=\nonumber\\
    &=&\frac{1}{\sqrt{C}}\left( \frac{1}{a}-\frac{2}{\sqrt{a}}-2\beta \frac{\ln a +1}{a} +1 +2\beta \right)\;.
\end{eqnarray}
The first-order Neumann series up to second order thus reads
\begin{eqnarray}\label{eq:1ser}
e^{(1)}(a)&=&\sum_{k=0}^2 a^k e_k^{(1)}(a)=
 \frac{1}{\sqrt{C}}\left( -\frac{a^{3/2}}{2}+(1+2\beta)a^2\right)\;.
\end{eqnarray} 

Now we can employ Eqs.~(\ref{eq:0ser}) and (\ref{eq:1ser}) and the \textit{true} coefficients of the expansion, and recalling the relations in Eq.~(\ref{eq:const}), we recover the inverse expansion rate for an Einstein-de Sitter universe:
\begin{eqnarray}\label{eq:eds_back}
  e(a)&=&\sum_{j=0}^1 \tilde{c}_j e^{(j)}(a)= \nonumber\\
   &=&\frac{2}{\sqrt{\alpha}}(1+2\beta) \sqrt{\alpha}a^2-
    2\sqrt{C}\frac{1}{\sqrt{C}}\left( -\frac{a^{3/2}}{2}+(1+2\beta)a^2\right)= \nonumber\\
     &=& a^{3/2} \;.
\end{eqnarray} 

\bibliographystyle{aa}
\bibliography{./master}

\end{document}